\documentclass[twocolumn,amsmath,amssymb,aps,floatfix, longbibliography,superscriptaddress]{revtex4-1}
\usepackage{graphicx}
\usepackage{dcolumn}
\usepackage{bm}
\usepackage{amssymb}

\begin{document}


\title{Constitutive relation for the system-spanning dynamically jammed region in response to impact of cornstarch and water suspensions}

\author{Rijan Maharjan}
\affiliation{Department of Mechanical Engineering and Materials Science, Yale University, New Haven, CT 06520}
\author{Shomeek Mukhopadhyay}
\affiliation{Department of Mechanical Engineering and Materials Science, Yale University, New Haven, CT 06520}
\author{Benjamin Allen}
\affiliation{Department of Mechanical Engineering and Materials Science, Yale University, New Haven, CT 06520}
\affiliation{School of Natural Sciences, University of California, Merced, California 95343}
\author{Tobias Storz}
\affiliation{Department of Mechanical Engineering and Materials Science, Yale University, New Haven, CT 06520}
\author{Eric Brown}
\affiliation{Department of Mechanical Engineering and Materials Science, Yale University, New Haven, CT 06520}
\affiliation{School of Natural Sciences, University of California, Merced, California 95343}

\date{\today}

\begin{abstract}
We experimentally characterize the impact response of concentrated suspensions consisting of cornstarch and water.  We observe that the suspensions support a large normal stress -- on the order of MPa -- with a delay after the impactor hits the suspension surface.  We show that neither the delay  nor the magnitude of the stress can yet be explained by either standard rheological models of shear thickening in terms of steady-state viscosities, or impact models based on added mass or other inertial effects.  The stress increase  occurs when a dynamically jammed region of the suspension  in front of the impactor propagates to the opposite boundary of the container, which can support large stresses when it spans between solid boundaries.  We  present a constitutive relation for impact rheology to relate the force on the impactor to its displacement.  This can be described  in terms of an effective modulus, but only after the delay required for the dynamically jammed region to span between solid boundaries.  Both the modulus and the delay are reported as a function of  impact velocity, fluid height, and weight fraction.  We report in a companion paper to this one on the structure of the dynamically jammed region when it spans between the impactor and the opposite boundary \cite{ASMMB17}.  In a direct follow-up, we  show that this constitutive model can be used to quantitatively predict, for example, the trajectory and penetration depth of the foot of a person  walking or running on cornstarch and water \cite{MAB18}.  
\end{abstract}

\maketitle

\label{sec:Intro}
 
Discontinuous Shear Thickening (DST) suspensions  exhibit a remarkable effect in which they behave like typical liquids at low shear rates, but when sheared faster, resistance to flow can increase discontinuously with shear rate \cite{Ba89,BJ14}.  DST suspensions can also exhibit solid-like properties such as cracking \cite{RMJKS13}. DST has been observed in a large variety of concentrated suspensions of hard, non-attractive particles, and is inferred to be a general feature of such suspensions \cite{Ba89, BJ12, BJ14,BFOZMBDJ10}. DST suspensions also support large stresses under impact, one example of which is the ability of a person to walk or run on the surface of a pool filled with a suspension of cornstarch and water  \cite{youtube_running, BJ14}.  The impact response of such fluids is of practical interest for impact protection gear because of their strong response during impact while remaining fluid and flexible otherwise \cite{LWW03,D3O}. The purpose of this paper is to obtain a constitutive relation that relates the force on an impactor to its displacement into a DST suspension.    A companion paper focuses on the internal structure of the suspension that leads to the strong impact response  \cite{ASMMB17}.   The development of a constitutive relation may aid in the development of materials for impact protection applications.

In steady-state rheology, shear thickening is defined by a range with a positive slope in the viscosity function $\eta(\dot\gamma) \equiv \tau/\dot\gamma$ as a function of shear rate, where $\tau$ is the shear stress and $\dot\gamma$ the shear rate in a steady-state shear flow.  The intent of such a constitutive relation is to predict flows with different forcing conditions, boundary conditions, and geometries.  The constitutive relation obtained from steady-state measurements indicates that suspensions of cornstarch and water can support steady shear and normal stresses up to $\sim 10^3$ Pa in a shear rate range where they are shear thickening, i.e.~before they become shear thinning (a negative slope of $\eta(\dot\gamma)$) at higher shear rates \cite{BJ12}. 
If we try to apply this  result from steady-state rheology to a person running on cornstarch and water, the predicted stress of $\sim10^3$ Pa is much less than needed for a person to be supported on the surface of the fluid,  based on a simple estimate of a person's weight distributed over the surface area of a foot ($\approx 4\cdot 10^4$ Pa).  Thus, the constitutive relation obtained from steady-state rheometer experiments fails to explain the strong response to impact.  It remains to be seen if our understanding of steady state DST can be extended to explain the strong impact response.

Recently an `added mass' model has been developed for impact response of dense suspensions, in which a `dynamically jammed' region forms ahead of the impactor in the fluid.  In this localized region, the suspension moves along with the impactor like a plug \cite{WJ12}.   The dynamically jammed region grows during the impact with a front which propagates away from the impactor \cite{WJ12,PJ14,HPJ16}.   There is a sharp velocity gradient at the front, which separates the dynamically jammed region from the surrounding fluid \cite{PJ14}. In a two-dimensional dry granular experiment the front velocity and width of the region with a velocity gradient both diverge at the same critical packing fraction as the viscosity curve of DST suspensions  \cite{WRVJ13}.  

In the  model for the added mass effect, the impact response of the suspension comes from an increasing suspension mass (i.e.~the `added mass') in the dynamically jammed region which moves with the impactor \cite{WJ12}.  This increasing mass slows down a free-falling impactor due to conservation of momentum.     
This model has been confirmed to quantitatively describe  the impact response of some high-speed projectiles into suspensions \cite{WJ12}.  However, to  significantly slow the impacting object by momentum conservation alone requires  large masses of fluid compared to the impacting object (or similarly, large depths of the fluid compared to the object's height).    The added mass model was not quantitatively applied to other impact response problems.  The regime of thin fluid layers where the added mass effect is weak is also particularly important for the related problem of impact protection applications where thin layers of protective material are desired \cite{DHNWW07,LWW03}.


When the dynamically jammed region reaches the boundary, the stress on the impactor increases beyond the added mass effect \cite{PJ14}.  However, it is not yet known how much more stress this boundary interaction can provide beyond the mass effect.  In particular, it is not known if this can provide more stress than steady state DST, or if it can explain the strong impact response cornstarch and water is known for. We report in a companion paper to this one on the structure of the dynamically jammed region when it spans between the impactor and the opposite boundary \cite{ASMMB17}.  We found that  the stress increase follows immediately after particle motion is observed at the boundary opposite impact.  We also observe dilation at this boundary in the same region where we find particle motion.  This observation is reminiscent of soils or dense granular materials.  It suggests a force transmission between particles along frictional contacts, as shear of a dense packing induces dilation as a result of particles pushing into and around each other.  This suggests the dynamically jammed structure could support a normal load  that is transmitted via frictional interactions across the system  when the dynamically jammed region spans from the impactor to a solid stationary boundary.  This assumes that the solid boundaries are much harder and have much more inertial mass than the fluid, so the solid boundary will not move or add to the mass  of the dynamically jammed region.  Instead, the relatively soft  dynamically jammed region deforms  as it is crushed between solid boundaries.  We hypothesize that the system-spanning dynamically jammed region could then temporarily support a load based on its effective stiffness, perhaps strong enough to support a person running on the surface.

To obtain a constitutive relation between force and displacement, we perform impact experiments where we measure the average stress response on the impactor  as a function of its displacement into the fluid.  The impactor is driven far enough  into  a suspension to see the dynamically jammed region interacting with the boundary, in contrast to previous experiments which probed mainly the response of the bulk \cite{WJ12,WRVJ13, PJ14, HPJ16}, but not so close to the boundary  to be affected by short-range boundary effects (i.e.~within $\approx 3$ mm) \cite{LSZ10}. Our experiments are at impact velocities  faster than quasistatic compression, so that  dynamically jammed fronts can exist, but at speeds slow enough that inertial effects are negligible  \cite{Bagnold54,CB13} (including added mass \cite{WJ12} and high Mach number effects \cite{LPWG10,POLMFH15, GKL17}). This intermediate velocity regime is where the steady-state DST transition occurs (typically at flow velocities $\stackrel{<}{_\sim} 10^{-2}$ m/s in rheometers \cite{MB17}), but surprisingly, systematic force measurements have not yet been reported in this regime as far as we know.

The remainder of the paper is organized as follows.   The materials and methods  of  suspension impact experiments are explained in Sections \ref{sec:materials} and \ref{sec:methods}, respectively.   Results of measurements of stress versus displacement  of the impactor are reported in Sec.~\ref{sec:stress}.  In Sec.~\ref{sec:onsetstress},  we show that the  stress response to impact  greatly exceeds that of previously known  steady-state rheology, added mass, or inertial scalings.  In Sec.~\ref{sec:onsetstress} we show that the strong response occurs as soon as the dynamically jammed region spans to the opposite boundary and the added mass stops propagating with the impactor.  In Sec.~\ref{sec:modulus}  we fit the stress response  to obtain an averaged constitutive model  for impact rheology.  This includes an effective compressive modulus of the  dynamically jammed region, and a delay before the modulus  comes into effect due to the time it takes for the dynamically jammed region to propagate and span between solid boundaries.  In a direct follow-up paper, we test this constitutive model  by showing it can quantitatively explain the ability of people to walk and run on the surface of cornstarch and water \cite{MAB18}.

\section{Materials}
\label{sec:materials}

The suspensions were made of  cornstarch  purchased from Carolina Biological Supply, and tap water near room temperature.  Measurements were made at a temperature of $22.0\pm0.6^{\circ}$C, where the uncertainty represents the standard deviation from day to day.  Weight fractions $\phi$ for cornstarch and water were measured as the weight of the cornstarch divided  by the total weight of cornstarch plus water.   Weight fractions of cornstarch and water are very sensitive to histories of temperature and humidity, so different data sets taken with relative humidity ranging from 8\% to 54\% in Sec.~\ref{sec:stress} are not directly comparable.    To avoid misinterpretation from false comparisons, we do not report weight fractions for different experiments in this section.  All samples nominally had weight fractions from 0.53 to 0.61, in a range where they all exhibited noticeable shear thickening when stirred by hand.   For data sets represented in a single plot, the experiments were taken over a short enough time period to have a humidity standard deviation of  6\%.  We report  measured weight fractions in Sec.~\ref{sec:modulus}  where systematic  weight-fraction-dependent measurements were done under constant relative humidity of $48\pm6\%$.   Under these conditions,  we found specific weight fractions such as the liquid-solid transition $\phi_c=0.609$ based on the onset of a yield stress to be reproducible within $\pm0.007$ \cite{MB17}.

We directly measured a density of $\rho=1200\pm20$ kg/m$^3$ for a suspension at $\phi=0.57$ based on the volume and weight of the suspension in a graduated cylinder.  If we extrapolate based on the fraction of cornstarch and water using the known density of water, the density of suspensions is not expected to deviate outside the uncertainty range for weight fractions from 0.51 to 0.63, covering our entire measurement range.

  Samples  were initially mixed on a vortex mixer until  no dry powder chunks were  observed. Before each impact measurement, samples were additionally stirred by slicing through them at least 5 times with a spatula at velocities low enough to avoid significant cracking of the suspension, and prevent large air bubbles from being trapped inside the suspension.   This additional stirring helps counter any systematic effects of settling  or compaction from previous experiments.  This procedure produced a level of reproducibility of $\pm 30\%$  in stress measurements,  equivalent to what we could achieve by making new samples  before each measurement.  If instead we did not stir between measurements or we forced air bubbles to get trapped in the suspension, the stress varied by around a factor of 2 from run to run.

\section{Methods}
\label{sec:methods}

\begin{figure}  
\centering
\vspace{0.2in}
{\includegraphics[width=0.35 \textwidth]{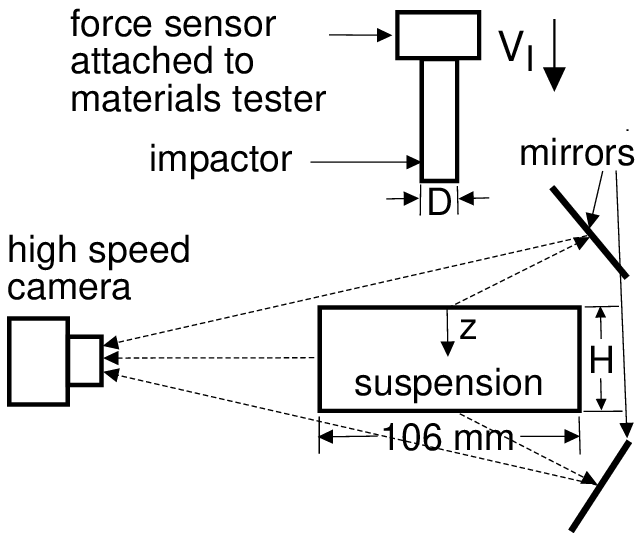}}
\caption{Schematic  of the experimental setup (side view). Measurement are made of the mean normal stress $\tau$ on the impactor as a function of impactor depth $z$ and impactor  velocity $V_I$. This can be done simultaneously with imaging of the top, bottom and side boundaries  of the suspension.}
\label{fig:apparatus}
\end{figure}

We performed experiments to visualize the  top, bottom, and side boundaries of the suspension to observe the dynamically jammed region, while simultaneously measuring forces  in response to impact, as shown in Fig.~\ref{fig:apparatus}.  The surface visualization results are reported in a companion paper \cite{ASMMB17}.  A cylindrical aluminum impactor of diameter $D = 12.7$ mm (unless otherwise noted) was pushed into a container with a square base of length $106$ mm, with the suspension filled to a height $H=42$ mm (unless otherwise noted).  These   dimensions are such that the  region  of interest below the impactor is far from the sidewalls of the container.  The impactor surface used for experiments reported in Sec.~\ref{sec:stress}  unintentionally had a slight wedge shape, which was angled at  $4^{\circ}$ relative to the surface.  The more quantitative experiments with controlled weight fraction reported in Sec.~\ref{sec:modulus} were done with a leveled impactor surface.  We are unaware of any effect of the misalignment on stress measurements, but to be safe we do not make direct comparison between those sections due to this misalignment and the differences in weight fractions.  We used an Instron E-1000 dynamic materials tester to push the impactor into the fluid at constant velocity $V_I$, while measuring the normal force on the impactor as a function of  depth $z$ from the free surface of the suspension (downward positive).   The nominal relative position resolution within each run is 1 $\mu$m.  We define $z=0$ and time $t=0$ at the top surface of the suspension, with an uncertainty of 0.5 mm.   The impactor started at a height typically $5.0 \pm 0.5$ mm above the suspension surface  and was pushed to a final position typically within $10\%$ of the bottom of the container.  While the impactor had a set point constant impact velocity $V_I$,  it had to accelerate at the beginning and end of the test. This resulted in a standard deviation of the velocity of the impactor for $z>0$ of 11\% for the data in Sec.~\ref{sec:stress}, and 5\% for the data in Sec.~\ref{sec:modulus}.  

Since the  force sensor was pushed against the impactor, the force measured by the sensor also included the force required to accelerate the mass of the impactor.   To correct for this, we performed a control experiment where the impactor moved through air (i.e.~with no sample).  The instantaneous acceleration is obtained from twice-differentiating the position signal.  We fit a linear relation between the normal force measured by the load cell and the instantaneous acceleration, where the proportionality corresponds to an effective mass of the impactor.  We subtracted out the corresponding force required to accelerate the impactor equal to the effective mass times the instantaneous acceleration from later force measurements. As a result of this inertial correction, the reported force results only from the force applied by the suspension due to impact, and the overall measurement noise is reduced significantly.  The force measurements are further calibrated by adding a small constant  so that the load cell reads zero force when nothing is pushing against the impactor.  This calibration is done separately for each measurement before the impactor hits the surface (i.e.~for $t<0$). This correction amounts to less than 1\% of the peak force measured. 

To reduce the remaining noise in stress measurements, inertia-corrected force data is smoothed over a range of $\pm0.5$ mm in $z$ to obtain a smoothed force $F$.   The average normal stress on the impactor is then given by $\tau=4 F/ \pi  D^{2}$.   We calculate the noise level $\sigma$ as the standard deviation of the smoothed stress $\tau$ for $t<0$, and only after the velocity stayed within $10\%$ of the set point velocity. 
In the event that the velocity threshold was not reached, we calculate the standard deviation over a minimum depth range of 1 mm instead.   We find $\sigma$ is roughly proportional to $V_I$ such that $\sigma=250$ Pa for $V_I=46$ mm/s, and reaches up to $\sigma = 3000$ Pa for our largest $V_I=584$ mm/s.  This increase is probably due to  transient accelerations, which become more significant as the momentum of the impactor increases.  These uncertainties amount to less than 0.14\% of our maximum stress signal.


\section{Stress response}
\label{sec:stress}

\begin{figure}
\centering
{\includegraphics[width=0.475 \textwidth]{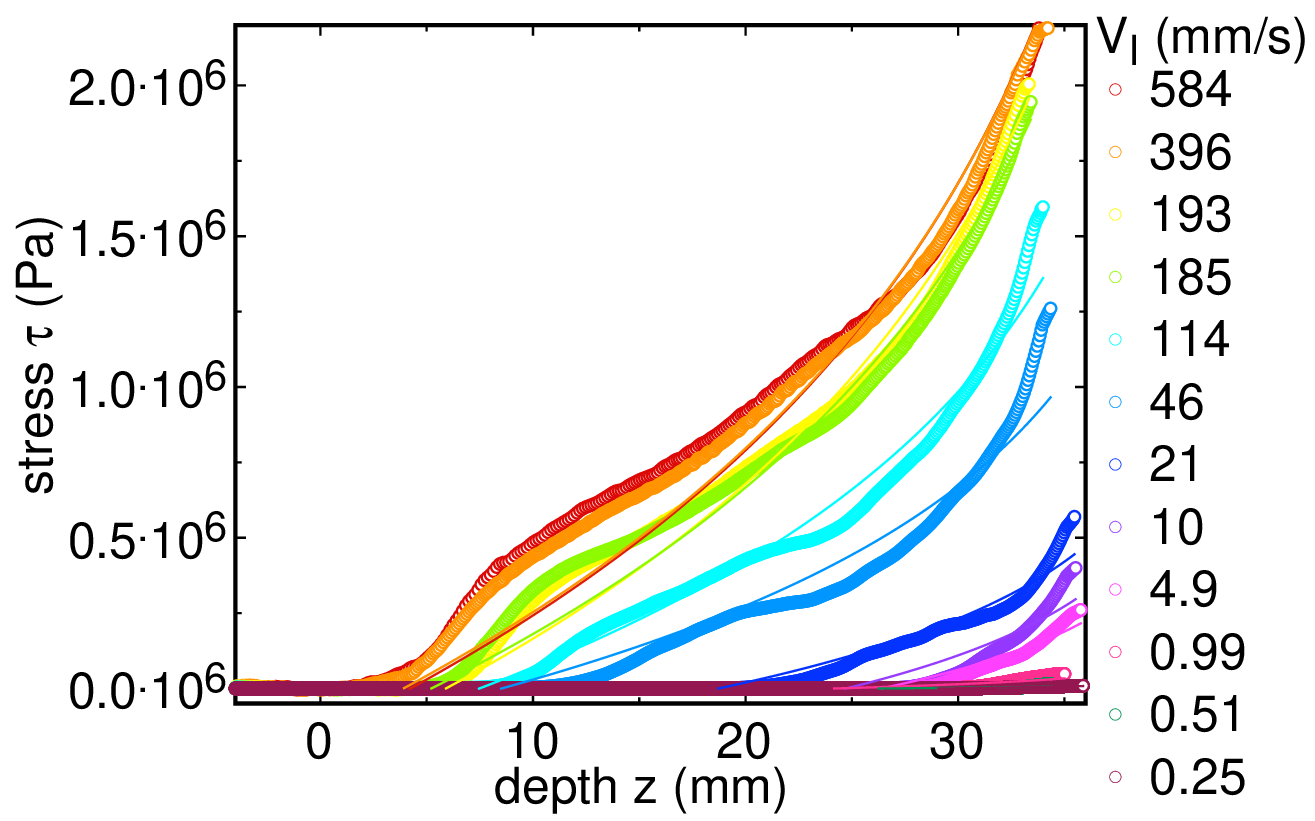}}
\caption {
(color online) The average  normal stress $\tau$ on the impactor vs.~depth $z$ of the impactor for several different impact velocities $V_I$ given in the legend (Upper curves correspond to larger $V_I$). Solid lines are fits to obtain a compressive modulus $E$ as described in Sec.~\ref{sec:modulus}.    In each case, a sharp increase in stress is observed, but with a delay after the point of impact ($z=0$). The scale of the stress reached is 3 orders of magnitude larger than can be explained by steady-state models for shear thickening, added mass or other inertial effects.
} 
\label{fig:stress_disp_all}
\end{figure}

\begin{figure}
{\includegraphics[width=0.475 \textwidth]{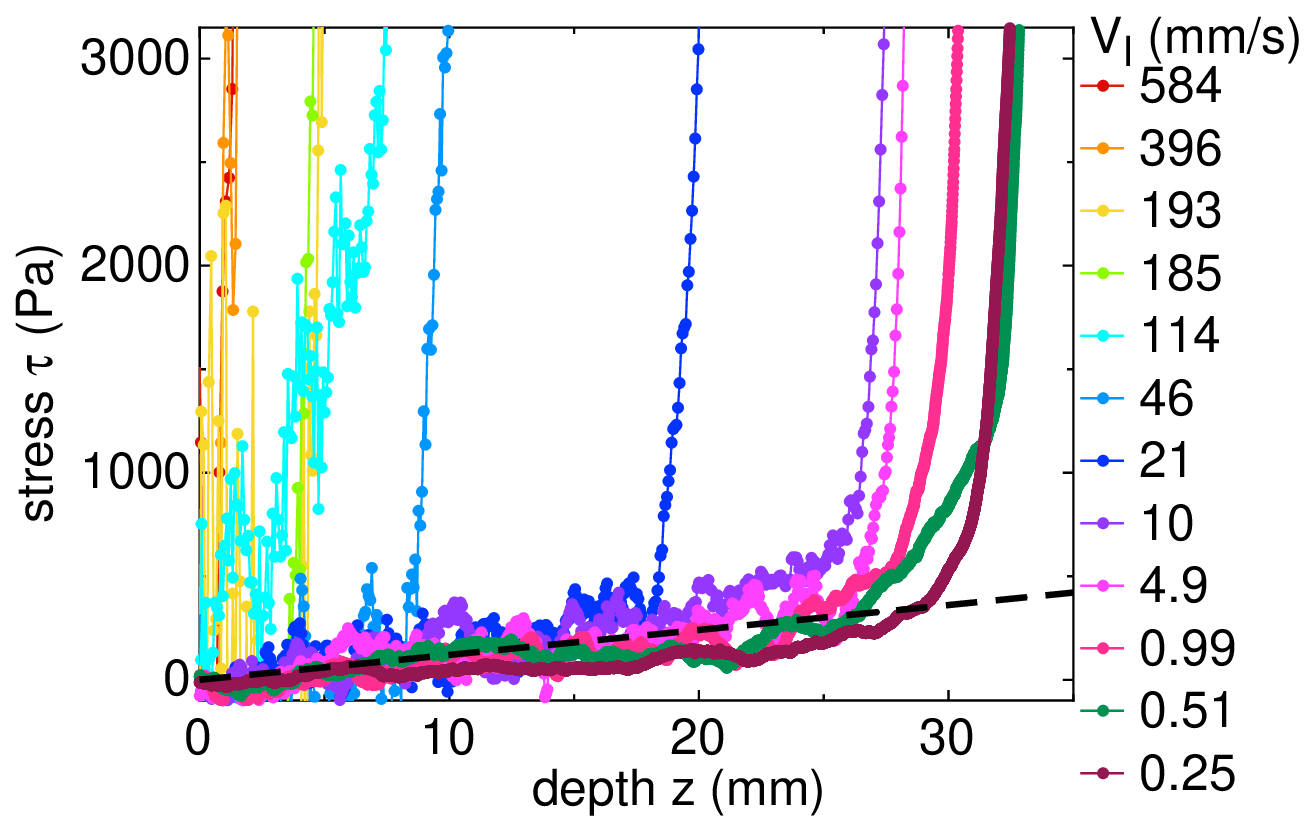}}
\caption { 
(color online) Data from Fig.~\ref{fig:stress_disp_all}, zoomed in to a smaller range of $\tau$ and connected by solid lines.  Dashed line: prediction of buoyant stress $\tau_b = \rho g z$.  Buoyancy can explain the weak  background stress response  at  the lower velocities $V_I \le 46$ mm/s. The  increase in stress  above the buoyant background is very sharp. 
}
\label{fig:buoyancy}
\end{figure}

To characterize the stress response to impact,  we performed measurements with the impactor moving into the fluid at constant velocity $V_I$. These data are plotted as normal stress $\tau$ vs.~depth $z$ in Fig.~\ref{fig:stress_disp_all} for several different values of $V_I$.  A striking feature is that for each curve, there appears to be a delay  between the point  where the impactor hits  the surface of the fluid (which defines depth $z=0$ and time $t=0$)   and the depth at which there is a noticeable non-zero stress. This  increase can be seen  more clearly to be a sharp   increase above a weak background when the same data is shown zoomed into a smaller  vertical scale in Fig.~\ref{fig:buoyancy}, indicating that the scale of the stress increases by several orders of magnitude at this sharp transition.   
The delay will be discussed further in Sec.~\ref{sec:onsetstress}.

 For  repetitions at any given  set of  experimental parameters, we observed a standard deviation of $30\%$ in $\tau$ as a typical run-to-run variation.    For $V_I < 10$ mm/s, we occasionally measured $\tau(z)$ curves where the sharp stress increase was not observed, i.e.~the stress did not increase beyond a scale of  $\sim 10^3$ Pa.    This non-reproducibility may be attributable to a large natural variation inherent in the mechanical response.  Such large variations have been observed for DST suspensions before, for example, even in steady-shear measurements the stress fluctuates by more than an order of magnitude over a  timescale of a second \cite{LDHH05}, which is longer than our experiments for $V_I \ge 46$ mm/s.  

To determine whether the deformation is elastic or plastic, we performed  extended experiments where after reaching the maximum  penetration depth, we set the applied force to zero.  If the deformation was mostly elastic, the impactor would have returned to near the suspension surface.   Instead, the impactor remained near its maximum penetration depth, and any upward motion was limited to a few percent of the penetration depth, indicative  of mostly plastic deformation.  In an alternative extended experiments we set the impactor  to move at velocity $V_I$ back to its initial position after reaching its maximum penetration depth.    In this case, the system did not retrace its original stress-deformation curve as an elastic system would, rather the stress dropped quickly to zero on its return trip, again indicating the deformation is mostly plastic.

\section{Comparison to existing models}
\label{sec:modelcomparison}

Before we present a full constitutive relation, we first compare our stress measurements to existing models for steady state rheology or impact response, which allows us to establish where the smaller background signal comes from, and identify what leads to the strong stress response.

\subsection{Buoyancy}
\label{sec:buoyancy}

 The dashed line in Fig.~\ref{fig:buoyancy} represents the stress due to buoyancy  on the impactor $\tau_b = \rho g z$ where $\rho=1200\pm20$ kg/m$^3$ and $g$ is  the acceleration of gravity. This  agrees with a good portion of the data,  in particular for $V_I \le 46$ mm/s and for small $z$  before the onset of the sharp stress increase.  This is consistent with impact experiments into  pure liquids in that velocity range, as well as for granular materials with possible small corrections for friction \cite{CB13}.  However,  This cannot explain data at larger $V_I$  or larger $z$.

\subsection{Steady state rheology models}

We next compare existing steady state rheology of DST to the stress in response to impact observed in Fig.~\ref{fig:stress_disp_all}.  The maximum shear and normal stress supported by cornstarch and water in the shear thickening range under steady shear is only $\sim 10^3$ Pa \cite{BJ12}.   This is still 3 orders of magnitude below the measured stress under impact.   Thus, we cannot explain the large magnitude of the stress in Fig.~\ref{fig:stress_disp_all} using the same viscosity function as steady state shear thickening in rheometer experiments.   
	
The local constitutive relation inferred from  steady-state shear experiments is dominated by a term where the shear stress and normal stress are related by an effective friction coefficient of order 1, mostly independent of the local shear rate \cite{LDHH05, BJ12, SMMD13, FMRKMLCHSI13, Heussinger13, RBH16, SGM17, CCNBSC17}.  That means the  local constitutive relation does not determine the magnitude of the shear and normal stresses in experiments.  Rather, the maximum stress reached in DST is a function of the boundary conditions, such that the  stress is limited by the stiffness of the weakest of the boundary or the particles in response to dilation \cite{BJ12}.  In rheometer experiments, typically the weakest boundary is the surface tension of the suspension-air interface, which  is what limits cornstarch and water to $\sim 10^3$  Pa.   In  cases where the stress is not limited by the suspension-air interface, the weakest stiffness could be soft walls \cite{BJ12} or the  particle stiffness \cite{WB09, BJ12} -- the latter case has been observed  in steady state flows in simulations with periodic boundary conditions \cite{OH11, SMMD13, SGM17}, but not in hard-particle experiments.   In our transient impact experiments, the strong stress response does require that the dynamically jammed region reach the boundary, so it is not a bulk response \cite{ASMMB17}.  Neither do the stresses propagate throughout the  entire system homogeneously to reach the suspension-air interface \cite{ASMMB17}, so the suspension-air interface is not what limits the maximum stress in  the impact response.  There could be a different stress-limiting mechanism in response to transient impact that applies  to the same local constitutive relation as steady-state flow, but this stress-limiting mechanism has not yet been identified.

This discrepancy in stress magnitudes between impact  and rheometer experiments holds even if we consider other known transient effects.  In rheometer experiments, the corresponding stress-strain curve during the transient (i.e.~before reaching steady-state) can evolve due to transient structure formation  \cite{GA80, FBOB09, BJ12}.  The corresponding stress-strain curve  exhibits a gradual  increase in stress without a delay, and the stress remains mostly lower than in the steady-state, with an occasional  overshoot of  the steady-state by about a factor of 2.  Strong fluctuations in stress have been observed around the steady state behavior \cite{LDH03, NN16},  although these have not been observed to exceed the order of magnitude of the steady-state average at the higher stress end of the shear thickening regime.

\subsection{Lubrication}
\label{sec:lubrication}

Viscous drag in the small lubricated gaps between particles is another possible source of stress that is often relevant in suspensions \cite{BB85}.   There is an upper bound on the  effective viscosity that can be obtained  from such lubrication forces before continuum hydrodynamics breaks down.  This occurs when the gap between solid surfaces is less than 2 liquid molecules thick \cite{VG88}.  The upper bound on the effective viscosity is $\eta = 9\eta_l a/8h \approx 40$ Pa$\cdot$s for cornstarch in water, where $\eta_l$  is the viscosity of the liquid phase, $a$ is the particle diameter, and $h$  is the liquid molecule diameter \cite{BJ14}.     The  corresponding upper bound on the stress is $\tau = \eta V_I/D$, which could reach only up to 2 kPa for our largest measured $V_I$. This is still 3 orders of magnitude too low to explain the strong impact response of the suspension, confirming that lubrication-based hydrodynamic mechanisms cannot explain the stress  increase above the background.  The higher stress and effective viscosity of the measured data suggests that the particles effectively collide  and more likely interact through effective solid-solid frictional interactions rather than lubrication.

\subsection{Inertial effects}

 At high impact velocities into fluids and granular materials, it is expected that inertial forces dominate, roughly corresponding to the force required to displace the inertial mass of fluid out of the way. The corresponding stress on the impacting object generally scales as $\tau_I \propto \rho V_I^2$,   regardless of the internal dissipation mechanics of the material. The proportionality coefficient varies from material to material from 0.1-4 \cite{Schlicting60, AMM57, UG10,CB13, BB17}, and can fluctuate around a mean value \cite{CKB12}. The largest coefficient of 4 yields  an estimate $\tau_I = 1700$ Pa  in our measured range of $V_I \le 600$ mm/s, still 3 orders of magnitude below the measured stress response, and not even  enough to hold a person's weight.  An extrapolation suggests that this would not overcome our maximum measured $\tau\approx 2\times 10^6$ Pa  until $V_I \stackrel{>}{_\sim} 2\times10^4$ mm/s.  High-speed impact studies of shear thickening fluids with $V_I \stackrel{>}{_\sim} 10^5$ mm/s have confirmed that inertial displacement determines the impact response \cite{POLMFH15},  consistent with this extrapolation.

\subsection{Added mass}
 
The model for the added mass effect is based on data for an object free-falling into a cornstarch and water suspension \cite{WJ12}.  Waituakaitus \& Jaeger calculated the effective stress on the impactor  $\tau_a$ as the change in momentum over time per unit area $A$, where $A=\pi D^2/4$.  We can modify their model to apply to our measurements for constant velocity impacts by calculating the momentum change as the product of a constant impact velocity $V_I$ and a mass increasing at at rate $dm_a/dt$.  The growth of the added mass $m_a$ over time was empirically fit by a frustrum shape based on the force response on free-falling objects \cite{WJ12}.  The added mass can be written as a function of penetration depth $z$ as
  
  \begin{equation} 
m_a= \frac{0.37\pi \rho }{3} \left(\frac{D}{2}+kz \right)^2 kz \ ,
\label{eqn:addedmass} 
\end{equation}

 \noindent where $\rho$ is the fluid density, $D$ is the impactor diameter, and $k$ is a free parameter  which represents the ratio of front velocity $V_F$ to impact velocity $V_I$ and depends on weight fraction $\phi$.  $dm_a/dt$ is obtained from the analytic derivative of Eq.~\ref{eqn:addedmass}
  
  \begin{equation} 
\frac{dm_a}{dt} = 0.37 \pi\rho kV_I \left(k^2z^2 + \frac{2kzD}{3} + \frac{D^2}{12}\right) \ .
\label{eqn:dma_dt} 
\end{equation}

\noindent Here we have used the identity $dz/dt= V_I$ for an added mass that moves along with the impactor.  This assumption is supported by the observation that the velocity of the dynamically jammed region  is the same as the  impactor velocity while they are in contact, corresponding to an  uncompressed dynamically jammed region  before it spans between solid boundaries \cite{PJ14}.  The stress on a constant-velocity impactor from the added mass effect is then

 \begin{equation} 
\tau_a = \frac{V_I}{A}\frac{dm_a}{dt}=\frac{0.37 \rho kV_I^2}{3} \left(1+\frac{8kz}{D}+\frac{12k^2 z^2}{D^2}\right)  \ .
\label{eqn:addedmass_constantvi} 
\end{equation}

\noindent While this expression can be compared directly to stress measurements, because $k$ is a free parameter, it does not indicate any upper bound on the strength of the added mass effect.

The added mass effect is in practice limited by the amount of fluid available in the suspension that can be converted to added mass. Once the dynamically jammed region reaches the boundary of the suspension opposite the impactor, the added mass can no longer propagate in the direction of the impactor, so no stress is expected from the added mass effect after this time.  For a constant velocity impact,  the ratio of front depth to impactor depth $z$ is the same as the ratio $k$ between front velocity to impactor velocity, so the maximum impactor depth for the added mass effect is $z=H/k$, which can be plugged in to produce an upper bound in Eq.~\ref{eqn:addedmass_constantvi}.   This does not eliminate $k$ from Eq.~\ref{eqn:addedmass_constantvi}, but it does indicate that when the peak stress is larger at larger $k$, it also drops off more quickly as the dynamically jammed region reaches the opposite boundary faster.   We can  come up with a $k$-independent bound on the added mass effect by realizing that the net momentum change on the impacting object  comes from conservation of momentum, which is limited by the mass available in the fluid as a function of $H$.  The net work  done on the impactor per unit area can be obtained from an integral of the stress-displacement curve of Eq.~\ref{eqn:addedmass_constantvi} up to depth $z=H/k$

\begin{equation}
 \frac{W_a}{A}=\int_0^{H/k} \tau_a dz = \frac{0.37\rho V_I^2 H}{3}\left(1+\frac{4H}{D}+\frac{4H^2}{D^2}\right) \ .
 \label{eqn:work_addedmass}
 \end{equation}
 
 \noindent   The value $W_a/A$  calculated for our  experiment parameter values contributes to less than 0.6\% of the integral of the measured $\tau(z)$ up to the peak stress  for $V_I=584$ mm/s shown in Fig.~\ref{fig:stress_disp_all}a, and $W_a/A$ is even smaller for lower $V_I$.   Since this result of Eq.~\ref{eqn:work_addedmass} is  independent of the only free parameter $k$, there is no longer any flexibility in the added mass model when comparing this work per unit area.   Furthermore,  this $k$-independence of the area under the $\tau(z)$ curve for the added mass mechanism is independent  of the specific form of the geometric factors of Eq.~\ref{eqn:addedmass_constantvi} that account for the geometry of the dynamically jammed region as prescribed by \cite{WJ12}.  Thus,  there is no way to achieve the  large stress response observed in parameter range of Fig.~\ref{fig:stress_disp_all} from the added mass mechanism, even by adjusting the parameter $k$ or the shape of the dynamically jammed region in the model.  According to Eq.~\ref{eqn:addedmass_constantvi}, the added mass effect is more relevant in impact response at larger $V_I$ and larger $H$.

\subsection{Onset of stress increase above the background}
\label{sec:onsetstress}

 So far, we have shown that the sharp increase in stress up to $\sim 10^6$ Pa in Fig.~\ref{fig:stress_disp_all} cannot be explained by any previously known scalings, including lubrication hydrodynamics, a confining stress from surface tension, and inertial mechanisms including the added mass effect.   Neither can the large scale of the stress be connected to steady-state rheology measurements which are traditionally assumed to describe flows in different geometries.  Previous results indicated that a stress increase beyond the added mass effect could be a result of  the dynamically jammed region reaching the opposite boundary from the impactor \cite{PJ14}.  Here we hypothesize that the stress increase we observe up to the MPa range is the result of the this dynamically jammed region spanning between opposite boundaries, which could then support a load according to its effective stiffness.  In this case, the delay depth $z_F$ beyond which the stress increases above the background should follow the relation $z_F=H/k$.   We test this hypothesis by  attempting to self-consistently fit the delay depth $z_F$ and the contribution to the background stress from the added mass effect, which provides values of $k$.
 
\begin{figure}
{\includegraphics[width=0.475 \textwidth]{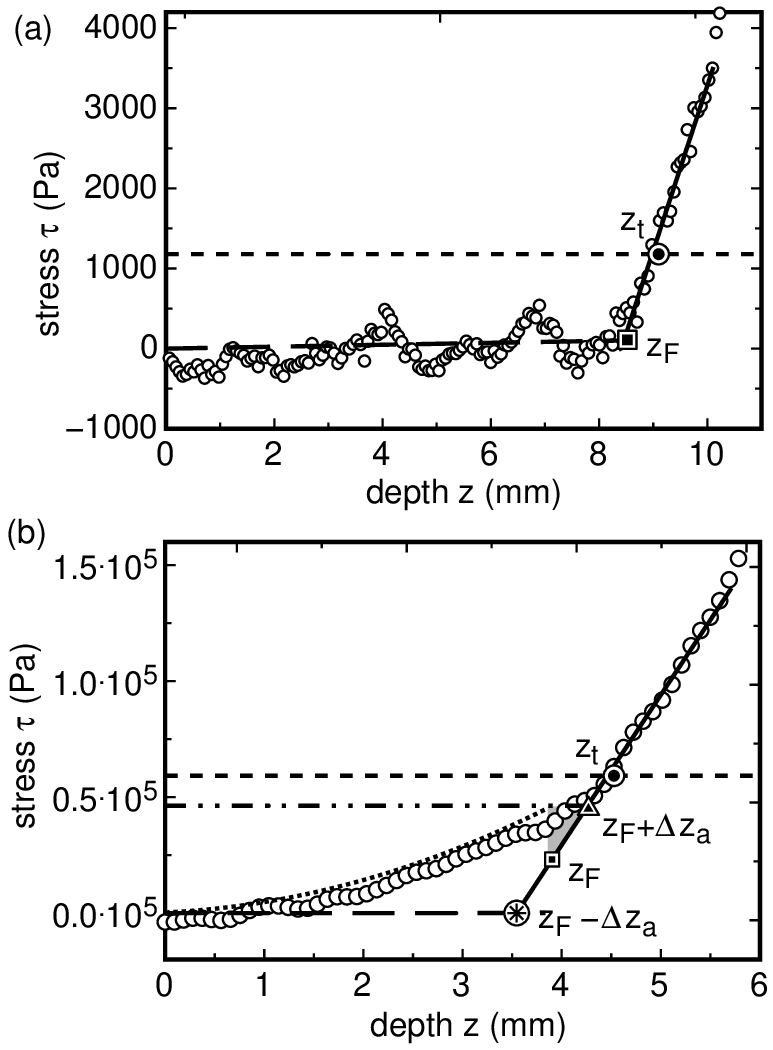}} 
\caption {Examples showing how we obtain the delay depth $z_F$, and correspondingly $k=H/z_F$.   (a) At $V_I=46$ mm/s, where the threshold stress $\tau_t$ (short-dashed line) from Eq.~\ref{eqn:threshold} is dominated by the noise threshold $5\sigma$.   An extrapolation of a linear fit (solid line) from above the threshold stress $\tau_t$  and displacement $z_t$ (partially filled circle) down to the background signal from buoyancy (long-dashed line) yields the onset depth $z_F$. (b) At $V_I=396$ mm/s, where the threshold stress $\tau_t$  is dominated by the added mass $\tau_a(z)$ (dotted line).  In this case, extrapolating the linear fit back to the background signal from both buoyancy and added mass (dashed-dotted line) yields an upper estimate of the delay depth $z_F+\Delta z_a$ (partially filled triangle),  and extrapolating the fit further to the background from only buoyancy yields a lower estimate of the delay depth $z_F-\Delta z_a$ (asterisk-filled circle).
}
\label{fig:example_fit_panel}
\end{figure}

A first estimate of the delay depth $z_F$ can be obtained as the depth where $\tau$ first exceeds a threshold from a sum of  contributions from buoyancy $\tau_b$,  the added mass $\tau_a$, and a noise threshold $5\sigma$, where $\sigma$ is the standard deviation of $\tau$ for $t<0$. This noise threshold is added so that rare fluctuations of the noise do not gives false results by exceeding the threshold prematurely.   The need to minimize fluctuations here motivated the smoothing  of force data explained in Sec.~\ref{sec:methods}.   Since we hypothesize that the  stress increase occurs when the dynamically jammed region first spans between solid boundaries, then we evaluate the added mass contribution from Eq.~\ref{eqn:addedmass_constantvi} when it has propagated across the system at depth $z=H/k$. The combined stress threshold  is given by 

\begin{equation}
\tau_t(z,k) = \rho g z + \frac{0.37 \rho kV_I^2}{3} \left(1+\frac{8H}{D}+\frac{12H^2}{D^2}\right) + 5\sigma \ .
\label{eqn:threshold}
\end{equation}
 
Since $\tau_a$ depends on $k$, and the range where this  expression is valid is up to depth $z_F=H/k$, we obtain $k$ and $z_F$ via an iterative fit process.   We start with an initial guess of $k=12$ \cite{WJ12} to calculate the threshold stress $\tau_t(z,k)$ from Eq.~\ref{eqn:threshold}.   We then compare the measured stress $\tau(z)$  to this threshold $\tau_t(z)$ to obtain a depth $z_t$  as the first estimate of the delay depth where the threshold $\tau_t$ is first exceeded.  Since this thresholding produces an overestimate, we attempt to extrapolate back to the intercept of the high-stress regime with the background signal.  We do this by fitting a local slope $d\tau/dz$ over the range $z_t < z < z_t + 1$ mm, or occasionally a larger range on the upper side if it resulted in a lower reduced $\chi^2$ (these larger fit ranges should give an equivalent fit with a smaller error).  At low $V_I < 46$ mm/s, where the background stress is dominated by bouyancy (Sec.~\ref{sec:buoyancy}),  the added mass contribution to the stress is small compared to $\tau_b$.  In these cases  we  linearly extrapolate the slope $d\tau/dz$ down to the  background from  buoyancy $\tau_b=\rho gz$ to obtain $z_F$ as shown, for example,  in Fig.~\ref{fig:example_fit_panel}a. This extrapolation helps minimize any errors introduced by the noise contribution to the threshold.  Once a value of $z_F$ was obtained, we then adjusted $k=H/z_F$ to input into Eq.~\ref{eqn:threshold} and iterated the process until the values of $k$ and $z_F$ were self consistent.    

For $V_I> 46$ mm/s, the background stress can be described by the added mass effect.  However, since the detailed contribution of the added mass effect to the stress for $z>z_F$ is not yet well-characterized, we had to consider reasonable upper and lower bounds.  The fit of $d\tau/dz$ was extrapolated back to two different values to act as bounds on $z_F$ as shown in Fig.~\ref{fig:example_fit_panel}b.   As an upper bound,  the fit was extrapolated to a background given by  added mass and buoyancy effects to obtain  $z_F + \Delta z_a = z_t-5\sigma/(d\tau/dz)$, corresponding to the added mass contribution to stress remaining constant for $z > z_F$ at the peak value reached.  As  a lower bound,  the fit was extrapolated to a background only from buoyancy $\tau_b$ to obtain $z_F -\Delta z_a = z_t -[5\sigma+\tau_a(z=H/k)]/(d\tau/dz)$,  corresponding to  the added mass contribution going to zero for $z>z_F$.   Previous  measurements  indicate  the actual response is somewhere in the range $z_F \pm \Delta z_a$ \cite{PJ14} where it appears that a small  remaining added mass effect for $z>z_F$ is enough to keep the total stress $\tau$ from decaying before the boundary contribution exceeds the added mass effect. This contribution would correspond to the area in the shaded triangle in Fig.~\ref{fig:example_fit_panel}b.  Our best estimate of the delay depth $z_F$ is then taken as the average of these two extrapolated bounds, with an error due  to the added mass effect of  $\Delta z_a = 0.5\tau_a(z=H/k)/(d\tau/dz)$.   Once a value of $z_F$ was obtained, we adjusted $k=H/z_F$ and iterated the process until the values of $k$ and $z_F$ were self consistent.    

These corrections and errors are small factors for most of the data presented in this paper.  The extrapolation from the first estimate $z_t$ to the final value for $z_F$  on average results in a small correction of 18\%. The error $\Delta z_a$ from the added mass effect  is on average  $2\%$, and always less than 10\% of $z_F$,  except for our largest  combination of $V_I=584$ mm/s and $H=200$ mm (not shown in Fig.~\ref{fig:stress_disp_all}), where $\Delta z_a = 0.81 z_F$, corresponding roughly to the transition where the added mass effect becomes the dominant contribution to the total measured stress $\tau$. 
There is also an uncertainty on $z_F$ of 0.5 mm  due to the uncertainty on the initial position  of the impactor relative to the surface of the suspension,  which is usually dominant, and more so at small $H$ where this error is a  significant percentage of the penetration depth.  Finally, there is an  uncertainty on the fit  of $d\tau/dz$,  which is obtained by adjusting the input errors to obtain a reduced $\chi^2\approx1$ for the fit, which contributes  to an error on $z_F$ and $k$ when extrapolated to the background.  This error is only  significant for a few runs at the smallest $V_I$ where $z_F$ is close to $H$ and the stress signal is weak.


\begin{figure}
{\includegraphics[width=0.475 \textwidth]{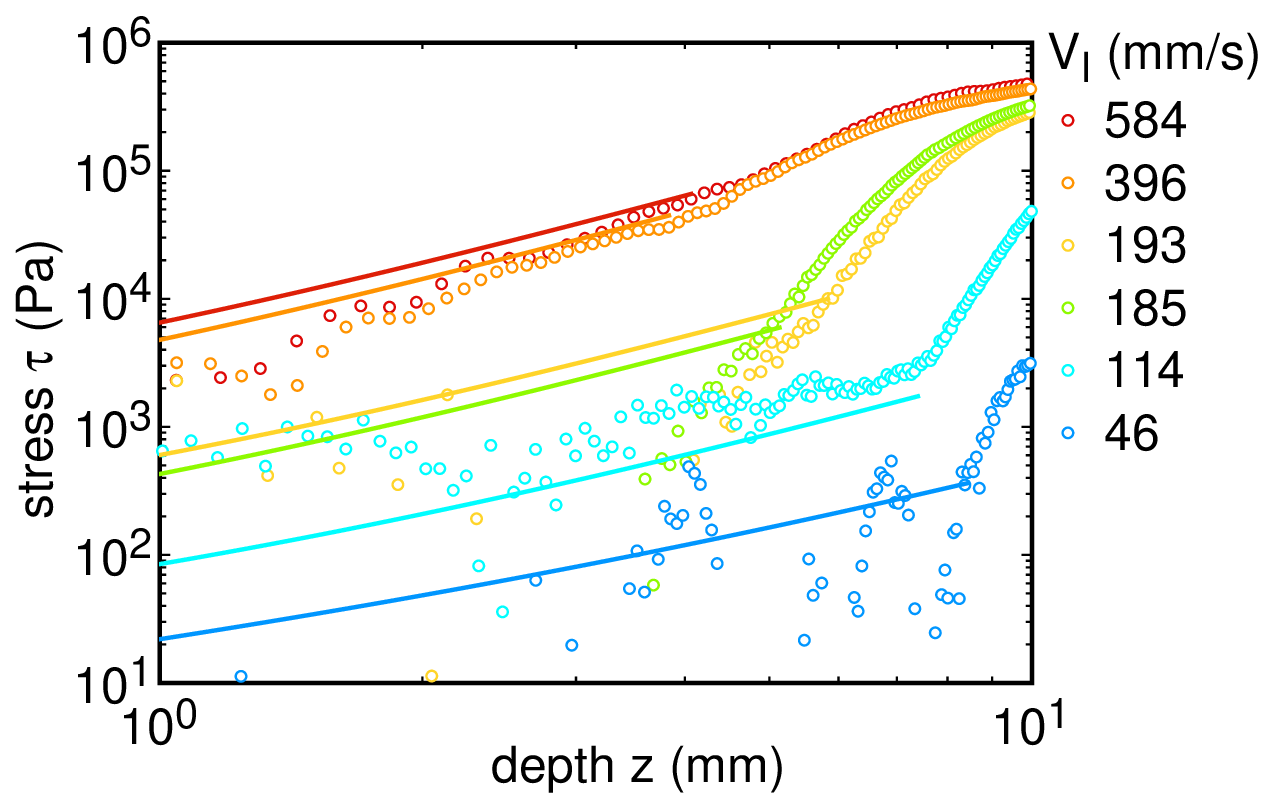}}
\caption {(color online)  The data from Fig.~\ref{fig:stress_disp_all}  zoomed into a smaller range of $z$ and on a log-log scale to focus on the added mass effect near onset for our largest $V_I$.  Solid lines:  added mass prediction from Eq.~\ref{eqn:addedmass_constantvi}, where $k$ is obtained as shown in Fig.~\ref{fig:example_fit_panel}.   The fits use the same color code  as the data.   The added mass effect is  consistent with the weak stress response  we observed before onset at  the highest velocities $V_I \ge 46$ mm/s. This consistency confirms that the sharp stress increase occurs when the front of the dynamically jammed region propagates to the opposite boundary.
} 
\label{fig:addedmass_constantvi}
\end{figure}

We next test whether this  iterative method to obtain the velocity ratio $k=H/z_F$  illustrated in Fig.~\ref{fig:example_fit_panel} self-consistently models  the added mass effect.  Figure \ref{fig:addedmass_constantvi} show data replotted from Fig.~\ref{fig:stress_disp_all}, zoomed into a smaller range of $z$ and on a log-log scale for our largest $V_I$  to see  the added mass contribution to the stress.  We plot the predicted contribution from the added mass effect (Eq.~\ref{eqn:addedmass_constantvi}) in Fig.~\ref{fig:addedmass_constantvi} for  each of our  $V_I\ge 46$ mm/s, using the value of $k=H/z_F$ obtained from  the iterative method  illustrated in Fig.~\ref{fig:example_fit_panel}, and plotting only up to $z_F$ in each case.  It is seen that in most cases the prediction  captures the scaling and magnitude the background stress,  although there is a lot of variation in the data and the measurements at $V_I=185$ mm/s and $V_I=193$ mm/s fall well below the predicted added mass effect.  Since the added mass contribution is less than 0.5\% of the total stress measured, and the difference between the data and the model is on the same order as the measurement resolution of 1000 Pa for these data sets, these differences may be a limitation of the measurement resolution combined with the natural variability of the stress from run to run.  Note that these predictions were made by fitting the onset of stress increase, not by fitting the added mass contribution to the stress directly.   Overall, this agreement confirms that,  within our limited resolution, the model of Waitukaitus \& Jaeger \cite{WJ12} describes the  contribution of the added mass effect to the background observed here for $V_I\ge 46$ mm/s.    This consistency check also confirms the hypothesis that the sharp stress increase occurs when the front of the added mass region reaches the solid bottom boundary.



\section{Constitutive model}
\label{sec:modulus}


In this section we obtain a constitutive relation that  quantitatively relates the average stress on the impactor to its displacement.  We base it on the observation in Fig.~\ref{fig:stress_disp_all} that the relationship between stress and strain or displacement is roughly linear, suggesting we can characterize the stress response after a delay with an effective compressive modulus of the system-spanning dynamically jammed region.  Mechanically, the dynamically jammed region could be much like a transient version of a soil or jammed granular material, where force would be transmitted across the system along effectively frictional contacts between particles, which is suggested by the observations of dilation  from visualization of the same experiments \cite{ASMMB17}.  The high stress level requires that these particle interactions are frictional rather than lubricated (as explained in Sec.~\ref{sec:lubrication}).  The deformation was also found to be mostly plastic (Sec.~\ref{sec:stress}), like a soil or granular material.

To obtain a relatively simple constitutive relation that describes the material response to impact, we report measurements of the velocity ratio $k$  between the front velocity and impact velocity which determines when  the  dynamically jammed region spans between solid boundaries at depth $z_F=H/k$, as well as an effective compressive modulus $E$ to approximately describe the  nearly linear stress increase observed afterward.  Both  measurements are reported over a range of impact velocities, suspension heights, and weight fractions.

\subsection{Height and impact velocity dependence of the velocity ratio $k$}

\begin{figure}
\centering
{\includegraphics[width=0.45 \textwidth]{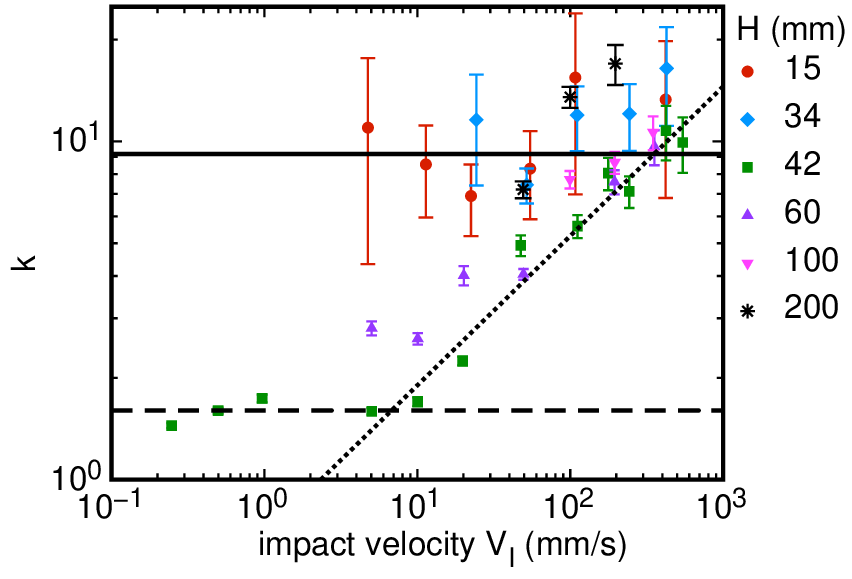}}
\caption{The ratio  $k$ between the front and impact velocities  as a function of impact velocity $V_I$ for different fluid heights $H$ shown in the legend.  Solid line:   constant fit to  the data for $V_I \ge 100$ mm/s. The collapse  for different $H$ confirms that the stress signal  propagates through the system  with a front velocity $V_F$ independent of $H$ for $V_I \ge 100$ mm/s.  Dashed line: constant fit for $V_I \le 10$ mm/s and $H=42$ mm. Dotted line: power law fit for $V_I \ge 10$ mm/s and $H=42$ mm.  The crossover of the dashed and dotted lines gives a minimum velocity $V_{min}$ for front propagation faster than the impactor velocity $V_I$.  For $V_I \le V_{min}$ mm/s, the values of $k$ do not collapse for different $H$, but are consistent  with a plug of aspect ratio near 1 moving along with the impactor. 
}
\label{fig:k_vi}
\end{figure}

The velocity ratio $k$ is calculated based on the ratio of travel distances of the front to the impactor as $k=H/z_F$, where $z_F$ was obtained from the fit method shown in Fig.~\ref{fig:example_fit_panel}.  These values of $k$  are shown in Fig.~\ref{fig:k_vi} as a function of impact velocity $V_I $ for different fluid heights $H$, at the same $\phi$ as the data in Fig.~\ref{fig:stress_disp_all}.  The errors plotted for $k$ are propagated from the errors on $z_F$.
For $V_I \ge 100$ mm/s, the data  scatter around a plateau value  for different $H$, suggesting a collapse of the data for different $H$ in this range.    Fitting a constant to $k$ in the range $V_I \ge 100$ mm/s for all $H$, with an input error of 30\% corresponding to the scatter, yields a plateau value of $k=9.2\pm0.8$  with a reduced $\chi^2 \approx 1$, confirming consistency with a plateau.   This $V_I$-independent $k$ is  similar to what was found by Waitukaitus \& Jaeger in this velocity range \cite{WJ12}.  The fact that $k$-values for different $H$ collapse, at least for $V_I \ge 100$ mm/s, confirms that in this range the delay is due to a signal propagation to the opposite boundary at constant velocity $V_F$ independent of $H$ in the bulk of the material,

For $V_I < 100$ mm/s in Fig.~\ref{fig:k_vi}, the values of $k$ do not  collapse at different $H$, indicating that in this range  the delay depends on something other than the bulk front  propagation velocity.   Specifically for $H=42$ mm, $k$ drops off to lower values at lower impact velocities, approaching close to $k=1$ for $V_I \le 10$ mm/s.  Physically, $k=1$ means that the  dynamically jammed region is not growing or propagating faster than the impactor.   An apparent $k$  slightly larger than 1 could be the result of a plug of jammed material in front of and moving at the same speed as the impactor, as suggested from the dead zone observed in particle tracking measurements \cite{ASMMB17}.  Our best fit of the plateau  in Fig.~\ref{fig:k_vi}  for $V_I \le 10$ mm/s and $H=42$ mm yields  $k=1.6$, which suggests a plug height of $H-H/k = 15$ mm.  This plug height is in between the impactor diameter (12.7 mm) and the width of the dead zone with no particle flow (20 mm) observed at the bottom boundary in these experiments \cite{ASMMB17}, corresponding to  a plug aspect ratio around 1, which is typical for granular flows.  

Assuming the values of $k$ at low $V_I$ are due to a plug moving with the impactor, we obtain a minimum velocity $V_{min}$ where  the front of the dynamically jammed region propagates faster than the impactor. We fit a constant to $k$ for $V_I \le 10$ mm/s, and a power law to $k$ for $V_I \ge 10$ mm/s at $H=42$ mm in Fig.~\ref{fig:k_vi}.  The crossover  of these two fits shown in Fig.~\ref{fig:k_vi} defines a minimum velocity $V_{min} =8\pm1$ mm/s,  where the errors on the fits are adjusted to obtain a reduced $\chi^2 \approx 1$.   

It is tempting to convert this critical velocity to a timescale $D/V_{min} = 1.8$ s and relate it to other timescales observed in DST suspensions. This is consistent with a stress relaxation time ranging from 0.01 to 2 s measured in rheometer experiments \cite{MB17}.   Such a connection could mean that the front propagation can occur because the local shear rate around the edge of the plug exceeds the relaxation rate, allowing the dynamically jammed region to grow.   It is also possible that $V_{min}/D$  is related to the critical shear rate $\dot\gamma_c$  from DST in rheometer measurements which varies over several orders of magnitude over the weight fraction range of DST.  To test either case would require measurements of relaxation time and  critical shear rate at the same weight fraction as impact experiments, measured over a range of weight fractions since the scaling of $\dot\gamma_c$ and the relaxation time are different in weight fraction.  Such a study is beyond the scope of this work.


\subsection{Weight fraction dependence of the velocity ratio $k$}

\begin{figure}
\centering
{\includegraphics[width=0.45 \textwidth]{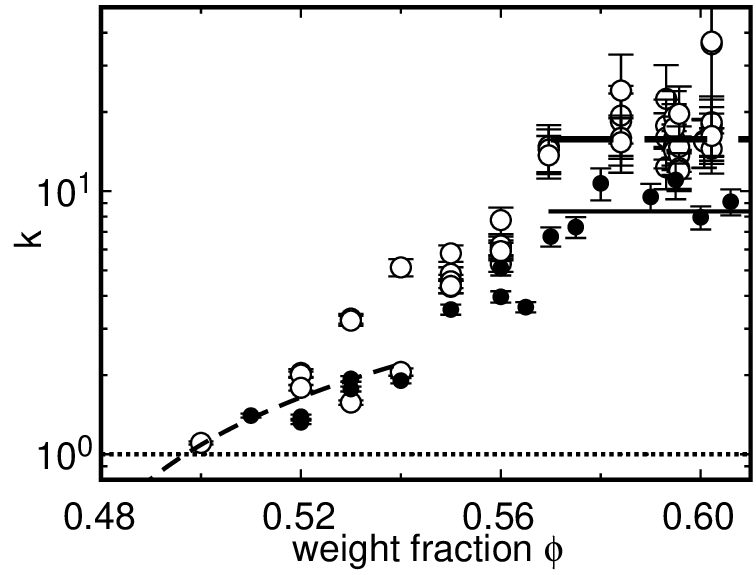}}
\caption{The ratio  $k$ between the front and impact velocities as a function of weight fraction $\phi$.  The two symbol sets  correspond to 2 different experimenters,  revealing a systematic difference for nominally similar procedures.    In each case, the value of $k$  reaches a plateau (indicated by the solid  and long dashed line fits) for  $0.57 \le \phi \le 0.61$,  up to the liquid-solid transition. Short dashed line:  a linear fit for $\phi \le 0.54$.  Extrapolating this fit to $k=1$ (dotted line) yields a  minimum  weight fraction $\phi_{min}=0.497$ for the existence of a stress  increase due to a dynamically jammed region spanning between solid boundaries.
} 
\label{fig:k_phi}
\end{figure}

The  values of the velocity ratio $k=H/z_F$ (where $z_F$ is obtained from the fit method shown in Fig.~\ref{fig:example_fit_panel}) are shown in Fig.~\ref{fig:k_phi} for different weight fractions $\phi$ with the impact velocity relatively fixed in the range 100 mm/s $\le V_I \le 400$ mm/s and $H = 42$ mm.  The value of $k$ increases with $\phi$, and it appears that the values from  $\phi \ge 0.57$ up to the liquid-solid transition could be consistent with a plateau.  The location of the liquid-solid transition at $\phi_c=0.61$ was  identified as the lowest weight fraction at which a non-zero yield stress is measured in rheometer experiments, using a portion of the same samples measured simultaneously to ensure the samples were at the same weight fraction \cite{MB17}.   We fit a constant $k$ to data in the range $0.57 \le \phi < 0.61$, which  yields a mean value $k=8.4 \pm 0.6$ with a reduced $\chi^2= 0.9$.  Instead fitting to a power law  in this range  yields an exponent consistent with zero, confirming the data are consistent with a constant value over this fit range $0.57 \le \phi \le 0.61$.   The lack of  divergence in $k(\phi)$ approaching the liquid-solid transition sharply contrasts with a quasi-2-dimensional dry disk model which has a divergence in $k$ as $\phi\rightarrow \phi_c$  \cite{WRVJ13}.   We do not know if the reason for this discrepancy is due to some consequence of adding the liquid, or a difference between 2 and 3 dimensional systems, or some other physics not included in the model.  


 For small $\phi$, $k$ approaches close to 1.   We could not resolve any stress increase above  the threshold $\tau_t$ in some cases for $\phi \le 0.54$, and in all cases for for $\phi \le 0.49$.   To determine if this  absence of signal  is an indication of a minimum $\phi$  for the existence of a dynamically jammed region, or a case of the signal dropping below the resolution, we fit  a linear function plus a constant to $k(\phi)$ for $\phi \le 0.54$, shown as the thin dashed line in Fig.~\ref{fig:k_phi}.    Error bars input to the fit were adjusted to a constant 27\%, corresponding to a run-to-run variation to obtain a reduced $\chi^2=1$.  Extrapolating this fit to $k=1$ yields a  minimum  weight fraction $\phi_{min}=0.497\pm 0.009$.  The  agreement of  the extrapolated $\phi_{min}$ with  the  consistent absence of  signal at lower weight fractions confirms that this is a minimum weight fraction for the existence of a stress  increase due to a dynamically jammed region spanning between solid boundaries.

 We found that  if the  measurements of $k(\phi)$ shown in Fig.~\ref{fig:k_phi} were repeated by another person with a nominally similar procedure with the same equipment,  then the values of $k$ shifted systematically.  In particular, the  plateau value of $k$ changed from 8.4 to 15.7 for the two data sets. This variation in $k$  for different experimenters suggests that the front propagation behavior is sensitive to details of the sample preparation and experiment which are not yet understood or well-controlled from experiment to experiment, such as the loading and stirring of the sample.  Such sensitivity may not be surprising in a system where even the run-to-run variation on repeated measurements is typically 20\%-30\% in most measured parameters.  Considering this wider range of values of $k$ we found, the value of $k=12$ previously obtained from experiments with an impactor in free-fall \cite{WJ12}  is within the  variation we observed in Fig.~\ref{fig:k_phi}.   The similarity of the results  suggests the model  for the propagation of the dynamically jammed region proposed by Waitukaitus et al.~\cite{WJ12} for free-falling objects also describes constant velocity impacts here.

\subsection{Method to obtain the modulus $E$}

To define a compressive modulus $E$ for a disordered solid with plastic deformation and no energetically preferred height for the dynamically jammed region, and with strains up to 0.9, a linear relation between stress and strain is not appropriate.  Instead we use a stress-strain relation for linear response that is continuously renormalized at each $z$-value:  $\tau = -E\ln[1-(z-z_F)/(H-z_F)]$. This is mathematically equivalent to calculating a modulus for a so-called `true strain' \cite{Shigleys}, although our stress-strain relation is still for the averaged impact response rather than a local stress-strain relation.  This approximates to a linear function $\tau\approx (z-z_F)E/(H-z_F)$ for small strain $(z-z_F)/(H-z_F)$ after the dynamically jammed region spans between solid boundaries when $z=z_F$. The logarithmic scaling accounts for the decreasing height of the dynamically jammed region over time.  In the range $z>z_F$, we expect the  added mass effect no longer contributes to the measured stress as the dynamically jammed region has stopped growing, so we characterize the total stress by the buoyancy term $\tau_b$ plus this modulus $E$ 

\begin{equation}
\tau = \rho gz -E\ln\left(1-\frac{z-z_F}{H-z_F}\right) \ .
\label{eqn:modulusfit}
\end{equation}

\noindent   We fit Eq.~\ref{eqn:modulusfit} to stress  measurements with only $E$ as a free parameter  and $z_F$  already determined by the fit methods shown in Fig.~\ref{fig:example_fit_panel}.  The fit range to find $E$ starts when the measured stress first exceeds the threshold stress $\tau_t$, and the fit ends at the maximum measured stress before we stopped the impactor.  Examples of fits to obtain $E$ for different $V_I$ are shown in Fig.~\ref{fig:stress_disp_all}.   While the detailed stress-displacement relation is more complex than Eq.~\ref{eqn:modulusfit}, this gives a simple two-parameter function for use in model predictions of the stress.  The error on this model given by the root-mean-square variation of data around these fits is 10\% of  the maximum stress for each curve in Fig.~\ref{fig:stress_disp_all}.  This level of error should be acceptable for  many purposes since the run-to-run variation on the overall magnitude of the stress is 30\%.

\subsection{Impact velocity dependence of the modulus $E$}

\begin{figure} 
\centering
\vspace{0.1in}
{\includegraphics[width=.475 \textwidth]{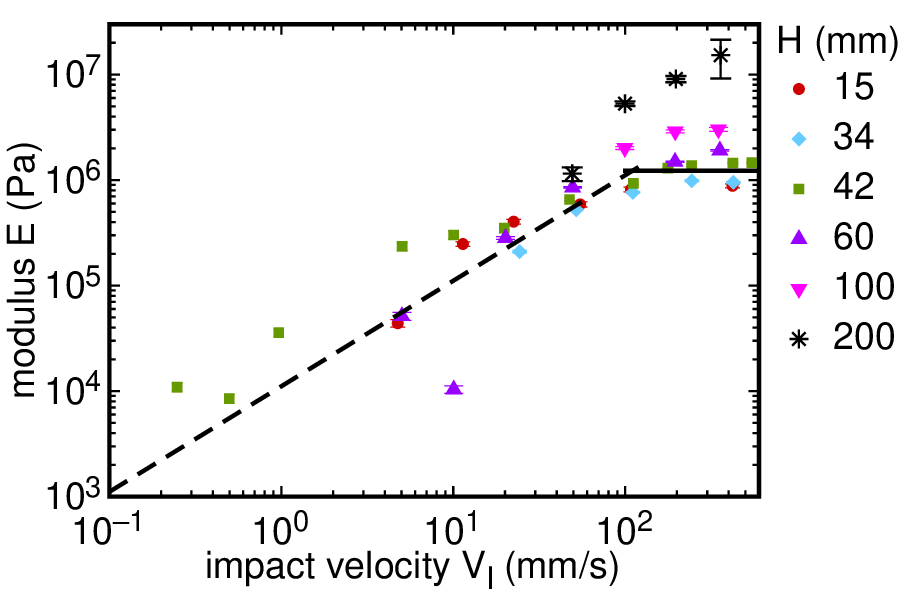}}
\vspace{0.1in}
\caption{The compressive modulus $E$  of the dynamically jammed region obtained from fits of $\tau(z)$.  Data are shown as a function of impact velocity $V_I$ for  different fluid heights $H$ as indicated in the legend.   A plateau is observed for $V_I\ge 100$ mm/s at each  $H< 200$ mm.  Solid line: constant fit to the data for $V_I \ge 100$ mm/s at $H = 42$ mm.  Dashed line: linear fit to the data for $V_I \le 120$ mm/s.
}
\label{fig:modulus_vi}
\end{figure}

  We measured stress-displacement curves for several impact velocities $V_I$, each at several different fluid heights $H$, at the same  weight fraction $\phi$ of the data in Fig.~\ref{fig:stress_disp_all}.   The fit values of $E$ are shown in Fig.~\ref{fig:modulus_vi}.  Errors shown for $E$ include the error on the fit, and an error propagated from the error on $z_F$ that determines the range of the fit.  The latter error can  be large if either $k$ is very small so the stress signal is small, or the added mass effect is so strong that the uncertainty propagated from $z_F$ leads to a large uncertainty in the fit range for $E$.   For larger $H$, some of the  experiments at lower $V_I$ never exhibited a stress signal above the background.  This is understood;  assuming $k$  remains independent of $H$, then the delay depth $z_F$ is expected to be larger than the machine-limited maximum penetration depth of the impactor of 55 mm.  
  
  A plateau in $E$ is observed for $V_I\ge 100$ mm/s  at each value of $H$ except for $H=200$ mm, consistent with the 14\% run-to-run standard deviation in $E$.   For example, the solid line in Fig.~\ref{fig:modulus_vi} shows constant fit to the data for $V_I \ge 100$ mm/s at $H = 42$ mm.  We note that the data at $H=200$ mm have the strongest added mass effect, in particular at $V_I=400$ mm/s the prediction for the added mass effect is 50\% of the total measured stress (i.e.~this is about the transition where the added mass effect overcomes the boundary effect in its contribution to the stress on the impactor).  The strong added mass effect likely influences the trends observed here at large $H$ and $V_I$. However, since we do not have a detailed model for the added mass effect after the collision of the dynamically jammed region with the boundary, it is difficult to account for it in detail here.   The plateau in $E(V_I)$ at large impact velocities $V_I$ is in contrast to what is expected from bulk rheology models where the shear stress remains linear in shear rate at stresses above the shear thickening regime \cite{WC14}, which would predict a linear relation between $E$ and $V_I$ in the limit of high velocity.

  For smaller $V_I$, the modulus $E$ scales approximately linearly with the impact velocity, as shown by the dashed line fit in Fig.~\ref{fig:modulus_vi}.  The non-systematic scatter in the data appears larger for $V_I \stackrel{<}{_\sim} 10$ mm/s, which is the same range where we found the occurrence of the sharp stress increase to be an unreliable  feature from run to run.  If we extrapolate the dashed line fit to lower velocities, the expected modulus becomes comparable to the background level due to buoyancy at $V_I=0.1$ mm/s

\subsection{Geometry dependence of the modulus $E$}

\begin{figure}
\centering
{\includegraphics[width=.475 \textwidth]{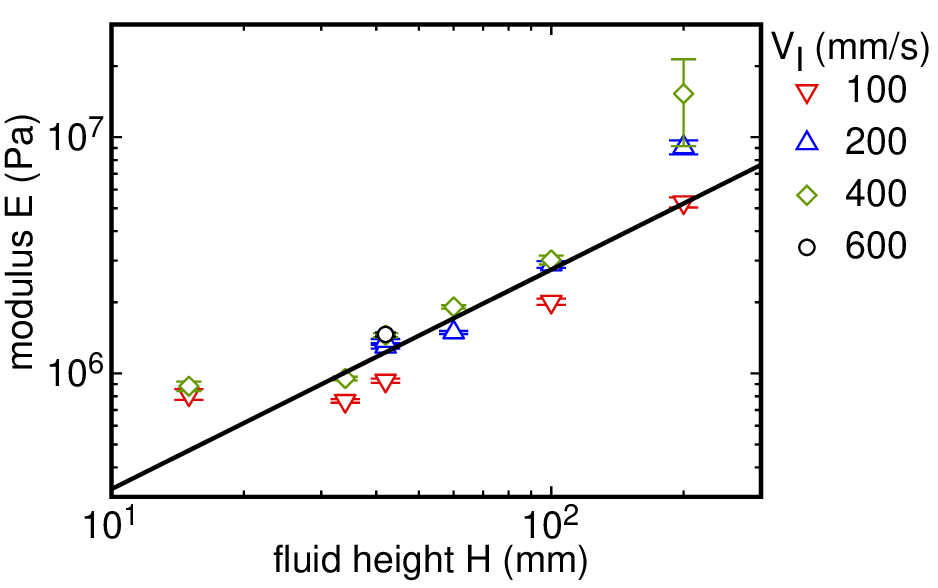}}
\caption{The modulus $E$  as a function of  fluid height $H$ for different $V_I$ as indicated in the legend. Solid line: power law fit to the data, yielding a best fit of $E\propto H^{0.93\pm 0.12}$.  $E$ is not independent of $H$ as would be the case for an intrinsic material modulus -- instead the trend is consistent with $E\propto H$, corresponding to a stiffness per unit area $d\tau/dz$ that  is independent of $H$.
} 
\label{fig:modulus_height}
\end{figure}

To characterize whether the stress response of the dynamically jammed region  scales like a bulk solid,  here we show how $E$ scales with the dimensions of the system.  In Fig.~\ref{fig:modulus_vi}, there  appears to be a systematic increase of $E$ with $H$. To quantify this, the values of the modulus $E$ in Fig.~\ref{fig:modulus_vi} are replotted as a function of fluid height $H$ in Fig.~\ref{fig:modulus_height} for a relatively narrow impact velocity range 100 mm/s $ \le V_I \le 600$ mm/s. The dashed line shows a power law fit  to these data where the input error was adjusted to 30\%  (about twice the typical run-to-run variation) to  obtain a reduced $\chi^2\approx1$.  This yielded  a power law exponent $0.93\pm0.12$. Since $E$ is found to vary with $H$, this indicates that $E$  is not an  intrinsic material property that is independent of system size.  Rather, the data are consistent with a linear scaling $E\propto H$.  Note that since $H-z_F =(1-1/k)H$, and $k$  is a constant in this parameter range, $E\propto H$ is equivalent to  $E\propto H-z_F$.  This means the fit in Fig.~\ref{fig:modulus_height} is consistent with a  height-independent stiffness per unit area $d\tau/dz = E/(H-z_F)$ of  the dynamically jammed region  in the limit of small strain  after it spans between solid boundaries (i.e.~for $z>z_F$).    If we allow for an the uncertainty in the scaling exponent of $d\tau/dz$ in $H$ of 0.1,  propagating this error only leads to a 30\% error in stress per decade of extrapolation in $H$. 
 This scaling implies that the mechanism that is setting the scale of the stress is coming from a bulk effect inside the suspension  that is  independent of the distance  from the impactor to the boundary, but only after the  dynamically jammed region spans between solid boundaries.

We also varied the diameter $D$ of the impactor (data not shown), and confirmed that over a range from $12.7 \le D \le 64$ mm, the measured modulus $E$ was consistent with no trend in $D$ over that range, within a standard deviation of 20\%, within the run-to-run variation.  This is  consistent with the hypothesis that the modulus $E$ is independent of size.  This also confirms that the shear force on the side of the impactor can be neglected in this range, as that would make a contribution to the modulus $E$ that scales as $1/D$.  However, for an impactor with $D=2.9$ mm, we found a modulus smaller by about a factor of 5, and a larger delay depth $z_F$ by about a factor of 3. This indicates the smaller impactor is in a parameter regime with very different scaling behavior than the rest of our data where $E$ is independent of $D$.  This regime at small $D$ was not studied in detail, and describing it  would require a constitutive relation dependent on two spatial dimensions to include a dependence on the impactor diameter $D$.  This is beyond the scope of this paper, as the current model describes a constitutive relation averaged over the horizontal plane and is only a function of one spatial dimension ($z$).

\subsection{Weight fraction dependence of the modulus $E$}

\begin{figure}
\centering
{\includegraphics[width=0.45 \textwidth]{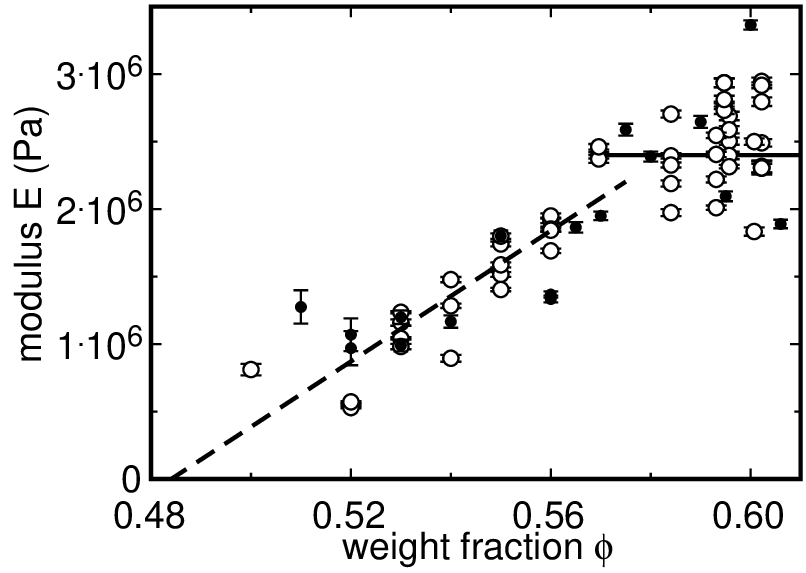}}
\caption{The modulus $E$ as a function of weight fraction $\phi$. The two symbol sets  correspond to 2 different experimenters.  Solid line:  constant fit to the data for $\phi\ge 0.57$. Dashed line: linear fit to the data for $\phi\le 0.57$.
} 
\label{fig:modulus_phi}
\end{figure}

We measured the modulus $E$ for various weight fractions $\phi$ with the impact velocity relatively fixed in the range 100 mm/s $\le V_I \le 400$ mm/s at $H=42$ mm, and show the results in Fig.~\ref{fig:modulus_phi}.   We measured weight fractions up to the liquid-solid transition at $\phi_c = 0.61$.  $E$ increases smoothly up to $\phi=0.57$, followed by a plateau in $E$.  In the range $0.57 \le \phi < 0.61$, a fit to a constant yields  $E=2.4$ MPa with a standard deviation of $0.3$ MPa, corresponding to $d\tau/dz = 64 \pm 9$ kPa/mm.   This happens to be the same range where $k(\phi)$ reaches a plateau (Fig.~\ref{fig:k_phi}).  At lower $\phi$, the modulus $E$ decreases.  A fit of $E(\phi)$ by a linear function plus a constant for $\phi \le0.57$ shown in Fig.~\ref{fig:modulus_phi} yields an intercept where $E=0$ at $\phi=0.484\pm0.006$.  This is consistent with $\phi_{min}$ where $k$ approaches the limiting value of 1 (seen in Fig.~\ref{fig:k_phi}) where the dynamically  jammed region does not propagate. Thus it seems  that the impact response gets weaker in terms of both $E$ and $k$ as the weight fraction is reduced to $\phi_{min}$, although at this point  the reason for this correspondence is not clear.

 We found that if the measurements of $E$ as  a function of $\phi$ shown in Fig.~\ref{fig:modulus_phi} were repeated by another  person with a nominally similar procedure,  the mean values of $E$ were relatively reproducible, well within each set's standard deviation of 14\%, as shown in Fig.~\ref{fig:modulus_phi}.  Thus, the values of $E$ seem less sensitive to the procedure than the values of $k$.

\section{Conclusions and discussion}

We observed that a suspension of cornstarch and water can support a large stress in response to impact with a delay after the impactor hits the suspension surface (Fig.~\ref{fig:stress_disp_all}). The strong impact response occurs when the dynamically jammed region responsible for the added mass effect spans between the impactor and the opposite solid boundary \cite{ASMMB17}.  The magnitude of this stress -- on the order of $10^6$ Pa -- cannot be explained by steady state rheology of DST, or impact models based on added mass or other inertial effects (Sec.~\ref{sec:stress}).  The background stress before this delay ($\sim 10^3$ Pa)  can be explained by a combination of buoyancy and added mass effects (Figs.~\ref{fig:buoyancy}-\ref{fig:addedmass_constantvi}).  


We used our measurements to obtain a relatively simple averaged constitutive rheology for impact response that  relates the force on the impactor to its displacement.  We characterized the impact response by a velocity ratio $k= V_F/V_I$ that determines  when the dynamically jammed region first spans to the boundary at impactor depth $z_F = H/k$ (Figs.~\ref{fig:k_vi} and \ref{fig:k_phi}). The system-spanning dynamically jammed region has an effective compressive modulus $E$ for $z>z_F$ (Figs.~\ref{fig:modulus_vi}, \ref{fig:modulus_height}, and \ref{fig:modulus_phi}).  The modulus $E=(H-z_F)d\tau/dz$ depends on the fluid height $H$, meaning that the dynamically jammed region does not have an intrinsic  material modulus like typical bulk materials.  Instead, we found  an intrinsic stiffness per unit area $d\tau/dz$ to be independent of $H$, which is  independent of the distance from the boundary, although this intrinsic response only exists after the  dynamically jammed region reaches the boundary.  We found the values of $d\tau/dz$ and $k$ to be constants over a wide range of parameters: impact velocities $100 \le V_I \le 600$ mm/s, weight fractions $0.57 \le \phi \le 0.61$ (up to the liquid-solid transition), suspension heights 15 mm $\le H \le 200$ mm, and impactor diameters 12.7 mm $\le D\le 64$ mm. In this range, we find $k=12\pm 4$ (Fig.~\ref{fig:k_phi}),  and  $d\tau/dz =64\pm 9$ kPa/mm (Fig.~\ref{fig:modulus_phi}),  where the error bars represent the large run-to-run standard deviation, including a variation for different experimenters in the case of $k$.   For  smaller $V_I$ and $\phi$, both $d\tau/dz$ and the velocity ratio $k$ drop off significantly.  We find a minimum velocity $V_{min}$ at a given weight fraction $\phi$ below  which the dynamically jammed region appears to act like a plug with aspect ratio near 1 that does not propagate faster than the impactor, although there is still a significant stress increase above the background.  We also found a minimum weight fraction  $\phi_{min} = 0.497\pm 0.007$ below which we did not observe any stress increase above the background at any velocity.


Despite a long-standing  expectation that the impact response of cornstarch and water  is related to shear thickening \cite{WB09},   there is still no quantitative observation that directly links the impact response of DST  suspensions to their steady-state rheology.  We observed dilation  in the dynamically jammed region,  which is a required part of the frustrated dilatancy mechanism of steady-state DST \cite{BJ12}.  We  also observed a weight fraction range of $(\phi_c - 0.04) < \phi < \phi_c$ where the impact response  reaches its maximum  strength,  in terms of modulus $E$ and the velocity ratio $k$.   The weight fraction range where $E(\phi)$  in Fig.~\ref{fig:modulus_phi} and $k(\phi)$ in Fig.~\ref{fig:k_phi} reach their maximum plateau values is the same weight fraction range where the stress-shear rate curve is discontinuous in  steady-state, rotation rate-controlled measurements  \cite{MB17}, indicating that the impact response and steady-state DST are strongest in the same weight-fraction range.  At this point,  these qualitative observations are the best evidence we have that the impact response and steady-state rheology might be connected, but this connection is tenuous at best.   Some other comparisons that could be made to steady-state DST include the existence of a  minimum weight fraction $\phi_{min} = 0.497 \pm 0.007$  for the strong impact response above the background level (Fig.~\ref{fig:k_phi}), and a minimum velocity $V_{min}$ for the dynamically jammed region to propagate faster than the impactor (Fig.~\ref{fig:k_vi}). It remains to be seen if the latter is related to the critical shear rate for  the onset of  shear thickening in steady-state DST, or a transient relaxation time \cite{MB17}. 
 
One major open question regards the physical origin of the large  stress scale on the order of $10^6$ Pa.   The observation that the modulus reaches a plateau as a function of both $V_I$ (Fig.~\ref{fig:modulus_vi}) and $\phi$ (Fig.~\ref{fig:modulus_phi}) is suggestive of a maximum or cutoff stress scale, analogous to the maximum stress in the shear thickening range of DST.  However, this scale has not yet been explained by any models.  In particular, this is 3 orders of magnitude larger than  the limit from surface tension at the suspension-air interface ($\sim 10^3$  Pa) that limits steady-state DST in rheometer measurements \cite{BJ12}.   In steady-state,  the stresses have time to distribute more uniformly throughout the suspension and are limited by the least stiff boundary.  In contrast, the  observation that the dynamically jammed structure is localized to a region below the impactor and does not need to reach the sidewalls of the system \cite{ASMMB17} suggests the normal stress transmitted along frictional interactions is limited by something inside the bulk of the suspension that exists during the transient.  

Recent work proposed that the stress scale on the order of MPa could come from the pore pressure: a pressure due to the liquid flowing between the pores of the granular packing as the granular packing rearranges \cite{JVF16}.  This model predicts a stress from pore pressure on the scale of $\tau_p\approx \eta_l \alpha\Delta\phi V_I L/\kappa$, where the viscosity of the interstitial liquid is $\eta_l=9\times10^{-4}$ Pa$\cdot$s, the permeability is $\kappa=(1-\phi^3)a^2/180\phi^2$, $\alpha$ is a dimensionless coefficient of order 1,  $L$ is the  width of the sheared region, and we interpret $\Delta\phi$ as the  change in weight fraction due to dilation from the initial value.  If we assume $\alpha=4$ \cite{JVF16}, an estimate for a typical value of $\Delta\phi\approx 0.01$ in a dilating suspension, and $L\approx 1.5$ cm based on the size of the portion of the dynamically jammed region that is sheared at $V_I=396$ mm/s \cite{ASMMB17}, then we obtain $\tau_p \approx 8$ MPa.
 This is promising that at  least pore pressure from dilation can produce a stress on the same order of magnitude as observed in Fig.~\ref{fig:stress_disp_all}, so the mechanism should be considered further.  However, the simple model was for a uniform fluid,  so could not even qualitatively predict propagation of the dynamically jammed region, and consequences of that such as the delay time or depth-dependence of the stress \cite{JVF16}.  A  complete model for the impact response should also be able to explain why the dynamically jammed region exists and propagates at all, its velocity, and the existence and values of  the minimum velocity $V_{min}$ and $\phi_{min}$ for front propagation.

Finally, the purpose of the constitutive relation is  to describe stresses, deformations, and flows of various impact-like phenomena; for examples with different forcing conditions, boundary conditions, flow geometries, and varying velocities.  In a follow up paper, we show that the constitutive relation can describe the ability of people to walk and run on the surface of cornstarch and water \cite{MAB18}.  To confirm that it is a generally useful constitutive relation requires it be tested on other transient impact-like phenomena as well.  There are several phenomena of shear thickening fluids which are not explained by steady-state viscosity functions, as least in the absence of time-dependent hysteresis terms.  One such phenomenon is the oscillation of the velocity of a sphere sinking in a  suspension, rather than monotonically approaching a terminal velocity as it would in a generalized Newtonian fluid \cite{KSLM11, KSM13, GBMP17}. It was argued that a repeated process of jamming and unjamming of something like  the dynamically jammed region could account for such oscillations  \cite{KSLM11, KSM13, GBMP17}.  Now we have an averaged constitutive model that includes such a process, along with a relaxation process to describe the unjamming \cite{MB17}.  Similarly, it was shown that the formation of stable holes in the surface of a vertically vibrated layer of a DST suspension could not be explained by a steady-state rheology in the absence of hysteresis in the constitutive relation $\tau(\dot\gamma)$ \cite{De10}. This apparent hysteresis appears to be time-dependent when the constitutive relation is put in terms of $\tau(\dot\gamma)$.  Alternatively, such hysteresis could come about for a history of increasing shear rate from the delay time we observed before the large stress increase.  For the history of decreasing shear rate, the the time dependence may come from the relaxation time of the dynamically jammed state \cite{MB17}.  Finally, the observation of objects bouncing off the surface of a DST suspension remains unexplained based on steady-state or added mass models  which are dissipative constitutive relations \cite{BJ14, WJ12}.  The system-spanning dynamically jammed region  can  in principle provide some energy storage in the modulus $E$ that could possibly explain the ability of impacting objects to bounce off the surface.   A detailed test of the application of the constitutive relation to  these and other problems is left open for future work.

\section{Acknowledgements}

We thank Abe Clark, Bob Behringer, Scott Waitukaitis, Ivo Peters, Heinrich Jaeger and Madhusudhan Venkadesan for discussions and for sharing their unpublished results. This work was supported by the NSF through grant DMR 1410157.

%

\end{document}